\documentclass[a4paper,pra,reprint,twocolumn,superscriptaddress]{revtex4}

\usepackage{ulem}
\usepackage{amssymb}
\usepackage{amsmath}
\usepackage{epsfig}
\usepackage{color}
\usepackage{graphics, graphicx}
\usepackage{bbold}
\usepackage{psfrag}
\usepackage{mathcomp}
\usepackage{amsmath}
\usepackage{amssymb}
\usepackage{mathrsfs}
\usepackage{subfigure}
\usepackage{verbatim}
\usepackage{xcolor}
\usepackage[colorlinks,citecolor=blue,urlcolor=blue]{hyperref}

\usepackage{mathrsfs}

\begin{document}

\title{Geometric phase of Wannier-Stark ladders in alkaline-earth(-like) atoms}
\author{De-Huan Cai}
\affiliation{CAS Key Laboratory of Quantum Information, University of Science and Technology of China, Hefei 230026, China}
\affiliation{CAS Center For Excellence in Quantum Information and Quantum Physics}
\author{Wei Yi}
\email{wyiz@ustc.edu.cn}
\affiliation{CAS Key Laboratory of Quantum Information, University of Science and Technology of China, Hefei 230026, China}
\affiliation{CAS Center For Excellence in Quantum Information and Quantum Physics}

\begin{abstract}
We discuss the geometric phase of Wanner-Stark ladders generated by periodically driven clock states in alkaline-earth(-like) atoms.
Using $^{171}$Yb atoms as a concrete example, we show that clock states driven by two detuned clock lasers can be mapped to two-band Wannier-Stark ladders, where dynamics of the system along the ladder is mapped to Bloch oscillations in a one-dimensional topological lattice. When the adiabatic condition is satisfied, the geometric phase accumulated in one period of the oscillation is quantized, and reveals the change of band topology as the laser parameters are tuned.
We show how the geometric phase can be experimentally detected through interference between different nuclear spin states. Our study sheds light on the engineering of exotic band structures in Floquet dynamics.
\end{abstract}

\maketitle

\section{Introduction}
An increasingly important tool for coherent quantum control, periodic driving is not only useful for the manipulation of atoms and spins through atom-photon couplings, but has also found applications in quantum simulation across a wide range of synthetic systems such as cold atoms~\cite{ETHcoldatom14,Weitenberg2016,yanbo}, trapped ions~\cite{Lukin17}, solid-spin systems~\cite{Lukin14,FRaman,caigeo}, photonics~\cite{KB+12,Rechtsman13,Cardano2017}, acoustics~\cite{Khanikaevnc}, and superconducting qubits~\cite{qubitqw}. Periodic driving gives rise to a wealth of interesting phenomena, such as topological charge pumping~\cite{thouless,pumpexp}, Floquet topological phases~\cite{CSPRA,Lindner11,Levin13,zhaizheng,zhouqi}, and dynamic topological constructions~\cite{Weitenberg17,Weitenberg1709,dqptqw,dynchern}, which have no counterparts in static systems. The synthetic degrees of freedom afforded by Floquet dressed states further offer intriguing possibilities of quantum control and quantum engineering, where these synthetic dimensions facilitate the observation of Floquet Raman transitions~\cite{FRaman} or the design of exotic lattice models~\cite{syn1,syn2,goldman}.

\begin{figure}[tbp]
  \centering
  \includegraphics[width=8cm]{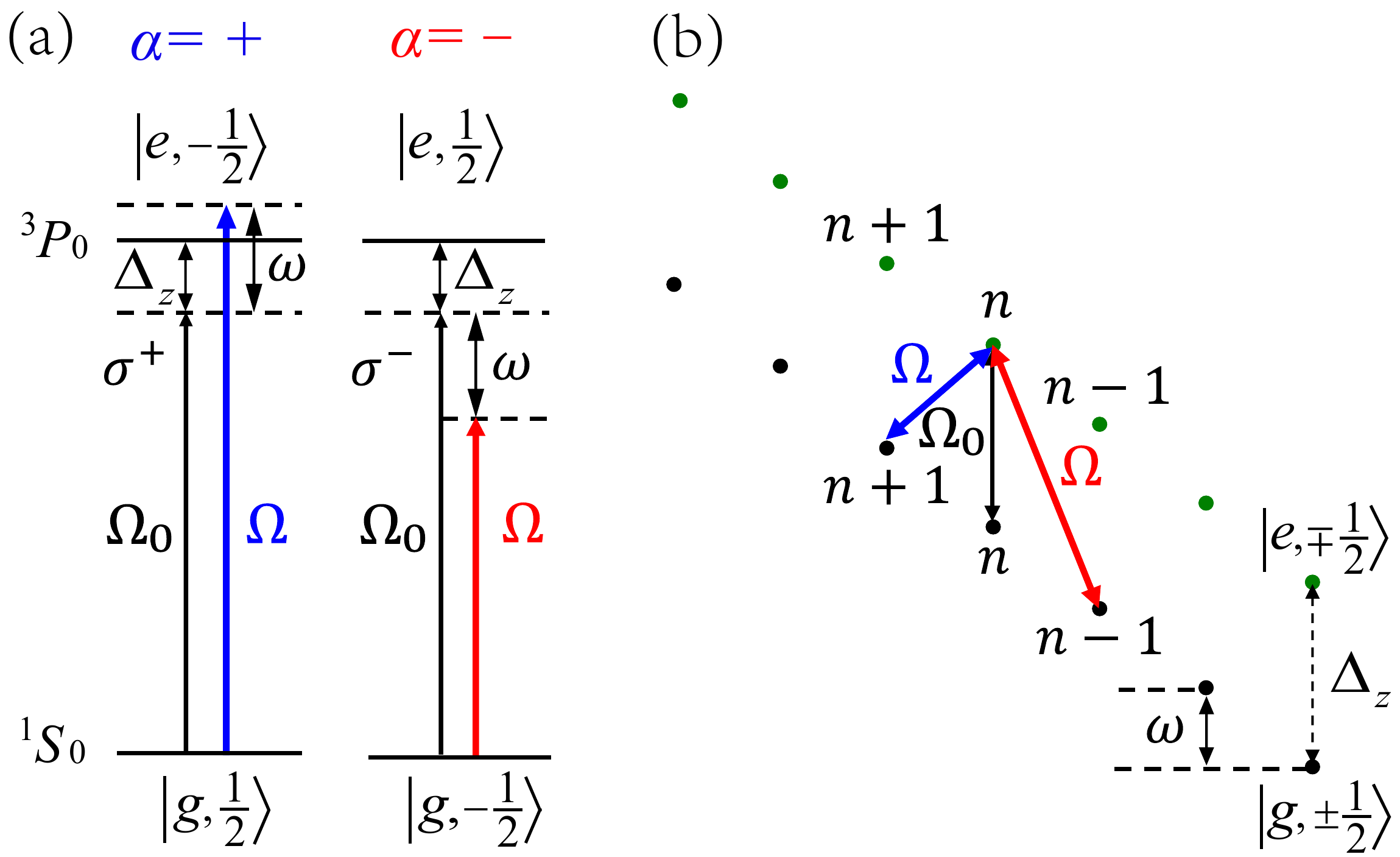}
  \caption{Schematic illustration of the coupling scheme and the resulting Wannier-Stark ladders. (a) Cross-coupled clock states of $^{171}$Yb atoms are divided into two groups: $\{|e,-\frac{1}{2}\rangle,|g,\frac{1}{2}\rangle\}$ ($\alpha=+$) and $\{|e,\frac{1}{2}\rangle,|g,-\frac{1}{2}\rangle\}$ ($\alpha=-$). For each group of hyperfine states, two detuned circularly polarized lasers couple the clock transition, with effective Rabi frequencies $\Omega_0$ and $\Omega$, respectively. The detunings $\Delta_z$ and $\omega$ are defined in the main text. (b) Two-band Wannier-Stark ladders of the Floquet Hamiltonian $H^{\alpha}_{\rm F}$, with different couplings for states in differnt groups: Blue and black arrows correspond to couplings for states in group $\alpha=+$; and red and black arrows correspond to couplings for states in group $\alpha=-$. Derivation and the explicit form of $H^{\alpha}_{\rm F}$ are given in the main text.}
  \label{fig:fig1}
\end{figure}

In this work, we show that Floquet dynamics in a periodically driven two-level system can simulate the Bloch oscillation in a one-dimensional lattice with topologically non-trivial bands. For concreteness, we use the clock states $^1S_0$ and $^3P_0$ of $^{171}$Yb atoms as an example, where the ground $^1S_0$ and the metastable $^3P_0$ manifolds are separated by an optical wavelength of $578$nm, and the electronic- and nuclear-spin degrees of freedom are decoupled~\cite{porsev1,porsev2,AE1,AE2}. For periodic driving, we consider a cross coupling of the clock states $|{}^1S_0,m_F=\pm \frac{1}{2}\rangle$ (labelled as $|g,\pm\frac{1}{2}\rangle$) and $|{}^3P_0,m_F=\mp \frac{1}{2}\rangle$ (labelled as $|e,\mp \frac{1}{2}\rangle$) with circularly polarized lasers, which divides the four clock states into two decoupled groups of two-level systems, as illustrated in Fig.~\ref{fig:fig1}. Under the rotating frame of the clock transition, the two-level system in each group is dictated by a periodically driven Hamiltonian, with the driving frequency $\omega$ given by the detuning between the coupling lasers.
We map the resulting Floquet dynamics for each two-level system to a two-band Wannier-Stark ladder~\cite{ws1}, which can be described as a tilted one-dimensional lattice in the synthetic Floquet dimension. Remarkably, these lattices possess topologically non-trivial bands, such that the Floquet dynamics can be understood as Bloch oscillations along topological lattices.
Under the adiabatic condition, which corresponds to a small laser detuning $\omega$, the non-trivial band topology is reflected in the quantized geometric phase the system acquires after one period of Bloch oscillation, which equals the Zak phase of the corresponding band~\cite{Zak,Bloch13}. In particular, as the driving parameters are tuned, the geometric phase undergoes an abrupt change, indicating a transition in the band topology. Away from the adiabatic regime with large laser detunings, the geometric phase is no longer quantized, with its behavior well-captured by time-dependent perturbations. Making use of the nuclear-spin degrees of freedom, we propose to detect geometric phases in the Floquet dynamics using interference measurements between the two groups of clock states. Our work explicitly demonstrates the potential of simulating exotic band structures and interesting topological phenomena in the synthetic dimensions available to Floquet dynamics.

The paper is organized as follows. In Sec.~II, we discuss the system setup and model Hamiltonian. We show how the Floquet dynamics is mapped to Wannier-Stark ladders and further to Bloch oscillations along topological lattices in Sec.~III. In Sec.~IV, we perform numerical calculations of the Zak phase in both the adiabatic and non-adiabatic regimes. We then discuss the experimental detection of our scheme in Sec.~V, before summarizing in Sec.~VI.

\section{Wanner-Stark Ladder as a tilted topological lattice}

We consider a cross-coupling scheme which divides the hyperfine states in the clock-state manifolds of $^{171}$Yb atoms into two groups of two-level systems.
As illustrated in Fig.~\ref{fig:fig1}(a), in the group labelled by $\alpha=\pm$, hyperfine states $|g,\pm\frac{1}{2}\rangle$ and $|e,\mp\frac{1}{2}\rangle$ are coupled by a pair of $\sigma^{\pm}$-polarized lasers. The general time-dependent Hamiltonian for each group is written as
\begin{align}
 H^{\alpha=\pm}_0 = \frac{\omega_r}{2}\sigma_z + \Omega_0\cos (\omega_0 t) \sigma_x + \Omega\cos[(\omega_0\pm\omega)t]\sigma_x,
\end{align}
where $\omega_r$ is the frequency of the clock transition, $\Omega_0$ and $\Omega$ are the Rabi frequencies of the coupling lasers, $\omega_0$ and $\omega_0\pm\omega$ are the frequencies of the coupling lasers with $\omega_0\gg\omega$, and $\sigma_i$ ($i=x,y,z$) are the Pauli operators associated with the two-level system in either group, with $\sigma_z=|e\rangle\langle e|-|g\rangle\langle g|$.

Applying the unitary transformation $U= \exp(\frac{i\omega_{0} t}{2}\sigma_z)$ and the rotating-wave approximation, we arrive at an effective Hamiltonian with periodically driven parameters
\begin{align}
H^{\alpha}_\text{eff}= \frac{\Delta_z}{2}\sigma_z + [\frac{\Omega_0}{2}+\frac{\Omega}{2}\cos(\omega t )]\sigma_x \pm \frac{\Omega}{2}\sin(\omega t)\sigma_y,\label{eq:effH}
\end{align}
where $\Delta_z = \omega_r - \omega_0$.
Following the standard derivation, the Floquet Hamiltonian $H^{\alpha}_F=H^{\alpha}_{\rm eff}-i\frac{\partial}{\partial t}$ can be written, in the second-quantized form, as
\begin{align}
&H^{\alpha}_\text{F} =-\sum_{n}n\omega(\hat{a}^{\dag}_{n,\alpha} \hat{a}_{n,\alpha} + \hat{b}^{\dag}_{n,\alpha} \hat{b}_{n,\alpha} )\nonumber\\
&+\frac{\Delta_z}{2} \sum_{n}(\hat{a}^{\dag}_{n,\alpha} \hat{a}_{n,\alpha} - \hat{b}^{\dag}_{n,\alpha} \hat{b}_{n,\alpha} )\nonumber\\
&+ \frac{\Omega_0}{2} \sum_{n}(\hat{a}^{\dag}_{n,\alpha} \hat{b}_{n,\alpha} + \hat{b}^{\dag}_{n,\alpha} \hat{a}_{n,\alpha} )\nonumber \\
&+ \frac{\Omega}{2} \sum_{n}(\hat{a}^{\dag}_{n,\alpha} \hat{b}_{n\mp 1,\alpha} + \hat{b}^{\dag}_{n\mp 1,\alpha} \hat{a}_{n,\alpha} ) , \label{eq:lattice}
\end{align}
where $n$ is the Floquet-band index, and $\hat{a}^{\dag}_{n,\alpha}$ ($\hat{b}^{\dag}_{n,\alpha}$) creates an atom in the state $e^{-in\omega t}|g,\pm\frac{1}{2}\rangle$ ($e^{-in\omega t}|e,\mp\frac{1}{2}\rangle$) for $\alpha=\pm$, respectively. As shown in Fig.~\ref{fig:fig1}(b), for each group of two-level system, Hamiltonian (\ref{eq:lattice}) is essentially a two-band Wannier-Stark ladder, with couplings indicated by blue and black arrows for states in group $\alpha=+$; and by red and black arrows for those in group $\alpha=-$.

Alternatively, Eq.~(\ref{eq:lattice}) can be rewritten as $H^{\alpha}_\text{F}:=H^{\alpha}_{\rm L}-\sum_{n}n\omega(\hat{a}^{\dag}_{n,\alpha} \hat{a}_{n,\alpha} + \hat{b}^{\dag}_{n,\alpha} \hat{b}_{n,\alpha} )$, which is identified
as a tilted Rice-Mele model, governed by $H^{\alpha}_{\rm L}$ in the synthetic Floquet dimension. Therefore, Floquet dynamics of the system can be mapped to Bloch oscillations along the tilted lattice. In particular, for $\Delta_z=0$, where $H^{\alpha}_\text{F}$ reduces to a tilted Su-Schrieffer-Heeger (SSH) model, geometric phases accumulated in the Floquet dynamics reflect the quantized Zak phases of the underlying topological bands when the adiabatic condition is satisfied.

\section{Geometric phase in Floquet dynamics}

To characterize geometric phases accumulated in the Floquet dynamics, we define creation operators in the synthetic-momentum space ($k$-space) by performing the Fourier transformation
$\hat{a}^{\dag}_{k,\alpha} = \frac{1}{\sqrt N} \sum_n e^{ink} \hat{a}^{\dag}_{n,\alpha} $ and $\hat{b}^{\dag}_{k,\alpha} = \frac{1}{\sqrt N} \sum_n e^{ink} \hat{b}^{\dag}_{n,\alpha} $,  where $k\in [0,2\pi)$. The effective lattice Hamiltonian $H^{\alpha}_{\rm L}$ can be written in the $k$-space as
\begin{align}
H^{\alpha}_{\rm L}=\sum_k\left(\begin{matrix} \hat{a}^{\dag}_{k,\alpha} & \hat{b}^{\dag}_{k,\alpha} \end{matrix}\right)H^{\alpha}_k \left(\begin{matrix} \hat{a}_{k,\alpha} \\ \hat{b}_{k,\alpha}\end{matrix}\right),
\end{align}
with
\begin{align}
H^{\alpha}_k = \frac{1}{2}\left[
   \begin{matrix}
     \Delta_z & \Omega_0+\Omega e^{\mp i k} \\
     \Omega_0+\Omega e^{\pm i k} & -\Delta_z \\
   \end{matrix}
 \right].
\end{align}
The quasienergy bands are the same for both $\alpha=\pm$, and are given by
\begin{align}
\epsilon_{k,i}= \pm\frac{1}{2}\sqrt {\Delta^2_z+(\Omega_0+\Omega\cos k)^2+\Omega^2\sin^2 k},
\end{align}
where $i=1,2$ is the band index.
For later reference, we write the creation operators for eigenstates of the quasienergy bands as
\begin{align}
\hat{c}^{\alpha\dag}_{k,i}=[\mu^{\alpha}_{k,i}]^T \left(\begin{matrix}a^{\dag}_{k,\alpha}\\b^{\dag}_{k,\alpha} \end{matrix}\right),
\end{align}
where
\begin{align}
\mu^{\alpha}_{k,1}&=\left(\begin{matrix} \cos\frac{\gamma_k}{2}\\ \sin \frac{\gamma_k}{2} e^{\pm i\theta_k}\end{matrix}\right),\\
\mu^{\alpha}_{k,2}&=\left(\begin{matrix} -\sin\frac{\gamma_k}{2}\\ \cos\frac{\gamma_k}{2}e^{\pm i\theta_k}\end{matrix}\right),
\end{align}
with $\tan\theta_k=\Omega\sin k/(\Omega_0+\Omega\cos k)$, $\tan\gamma_k=2|\epsilon_k|/\Delta_z$ and $|\epsilon_{k}|= \frac{1}{2}\sqrt {(\Omega_0+\Omega\cos k)^2+\Omega^2\sin^2 k}$.

The tilting of the lattice leads to an effective driving force, which leads to a Bloch oscillation along the lattice. For atoms initialized in the upper or lower band, they would adiabatically follow the corresponding quasienergy band, when $\omega$, or effectively, the driving force is small. In contrast, when $\omega$ becomes comparable with other parameters such as $\Omega$ or $\Omega_0$, Landau-Zener tunneling dominates, giving rise to the population of the other band. In the following, let us first focus on the adiabatic dynamics with small $\omega$.

When the adiabatic condition is satisfied, the time-dependent single-particle field operator can be written as
\begin{align}
 \hat{\Psi}^{\dag}_{\alpha} (t) = P_0e^{i\varphi^{\alpha}_{1}(t)}\hat{c}^{\alpha\dag}_{k,1} + Q_0e^{i\varphi^{\alpha}_{2}(t)}\hat{c}^{\alpha\dag}_{k,2},
\end{align}
where the dynamics is encoded in the evolution of the phase factors $\varphi^{\alpha}_{i}(t)$, while the initial condition is given by $P_0$ and $Q_0$. Applying the Heisenberg equation $\frac{d}{dt}\hat{\Psi}^\dag_\alpha (t)=\frac{i}{\hbar}[\hat{H}_\text{F}, \hat{\Psi}^\dag_\alpha (t)]$, and considering an adiabatic sweep of the synthetic momentum $k(t)=k_0-\omega t$ ($k_0$ is the initial synthetic momentum), we have
\begin{align}
\dot{\varphi}^{\alpha}_i (t) = \epsilon_{k(t),i} + i\omega\langle\mu^{\alpha}_{k(t),i}|\partial_k \mu^{\alpha}_{k(t),i}\rangle,
\end{align}
where $|\mu^{\alpha}_{k,i}\rangle:=\hat{c}^{\alpha\dag}_{k(t),i}|0\rangle$.

The key message here is that for atoms initialized within the same band but within different groups of clock states (with different $\alpha$ index), after one period of Bloch oscillation $T=2\pi/\omega$, they would accumulate the same dynamics phase, but the opposite Zak phase. Here the dynamic phase and the Zak phase associated with the $i$th band are respectively defined as
\begin{align}
\varphi_{{\rm d},i}&=-\int_0^T\epsilon_{k(t),i} dt,\\
\varphi_{{\rm z},i}&=i\int_0^T \langle\mu^{\alpha}_{k(t),i}|\partial_k \mu^{\alpha}_{k(t),i}\rangle.\label{eq:dz}
\end{align}
Our results thus suggest that, by probing the relative phase between clock states in different groups, we can extract Zak phase of the topological bands in the Floquet synthetic dimensions.

\begin{figure}[tbp]
  \centering
  \includegraphics[width=9cm]{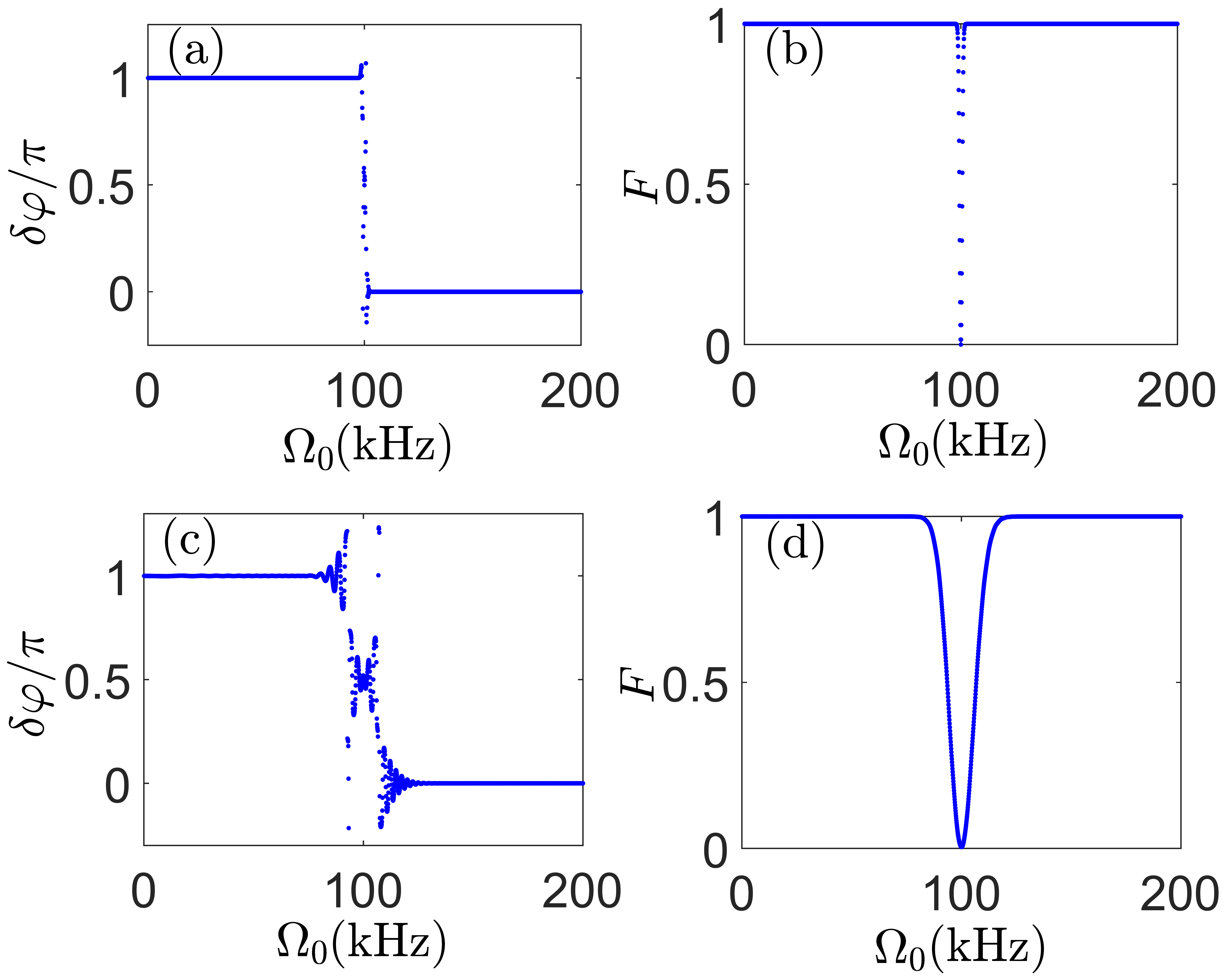}
  \caption{(a)(b) Numerically calculated $\delta\varphi$ and fidelity $F$ as functions of $\Omega_0$, for $\omega=0.01$kHz.
    (c)(d) Numerically calculated $\delta\varphi$ and $F$ for $\omega=1$kHz.
  For our calculations, we take $\Omega=100$kHz and $\Delta_z=0$. For both cases, we initialize the atom at $k_0=0$ in the upper band. Correspondingly, $\delta\varphi$ approaches the Zak phase of the upper band, which shows a topological transition near $\Omega_0=\Omega$. While the transition is reflected as a precipitous drop of the fidelity due to the closing of the band gap, with increasing $\omega$, the adiabatic condition begins to breakdown, as is clearly seen in (c)(d).
  }
    \label{fig:fig2}
\end{figure}

\section{Numerical simulation}

To confirm the conclusion above, we perform numerical simulations under typical experimental parameters of $^{171}$Yb. We initialize the system in the state $\frac{\sqrt{2}}{2}(\hat{c}^{+\dag}_{k=0,1}+\hat{c}^{-\dag}_{k=0,1})|0\rangle$, which corresponds to a superposition of eigenstates
at $k=0$ of the upper quasienergy bands of the two groups of clock states.
For our numerical simulation, we first set $\Delta_z=0$, and evolve the system under the Hamiltonian (\ref{eq:effH}) with $\omega\ll \Omega$ for one period of the Bloch oscillation $T$.
In Fig.~\ref{fig:fig2}(a), we show the phase difference $\delta\varphi=(\varphi^{+}_{1}-\varphi^{-}_{1})/2$ at the end of the time evolution, which
changes from $\pi$ to $0$ as $\Omega_0$ is tuned. This is consistent with the change of band topology of the underlying SSH model, where a topological phase transition occurs at $\Omega_0=\Omega$. It is also clear from Fig.~2 that, when $\omega$ is small, the Zak phase is quantized at $\pi$ and $0$ except for a very narrow region close to the phase transition point, where the gap between the upper and lower quasienergy bands becomes small and the adiabatic condition breaks down.
For larger $\omega$, the non-adiabatic region becomes wider, though the Zak phase is still quantized for parameters sufficiently away from the transition point [see Fig.~\ref{fig:fig2}(c)].

Such a picture is also reflected in the fidelity of the final state relative to the initial state, defined as $F=|\langle0|\Psi(0)\Psi^\dag(t)|0\rangle|$, which should approach unity under the adiabatic condition.
As shown in Fig.~\ref{fig:fig2}(b)(d), the fidelity rapidly drops to zero near the location where the band topology changes, suggesting the dominance of Landau-Zener tunneling close to the transition point.

\begin{figure}[tbp]
  \centering
  \includegraphics[width=9cm]{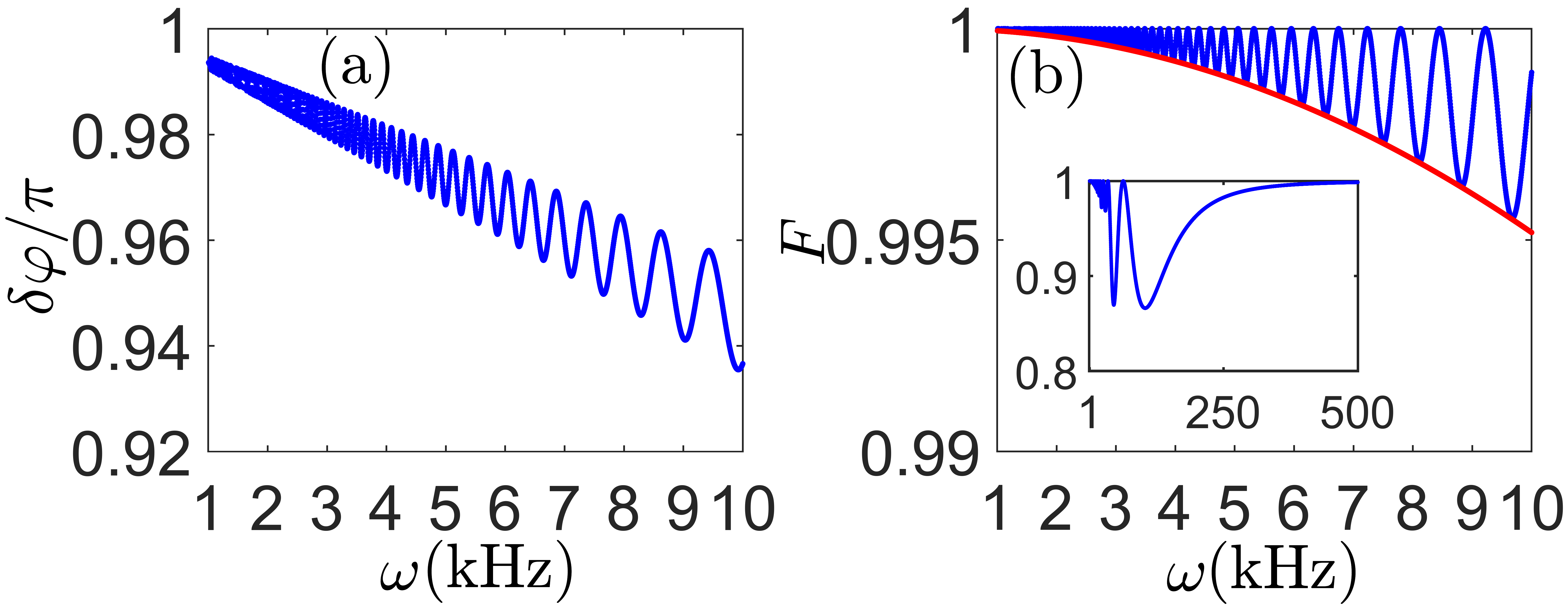}
  \caption{(a) $\delta\varphi$ and (b) $F$ as functions of $\omega$ for $\Omega_0=20$kHz. With increasing $\omega$, $\delta\varphi$ deviates from the Zak phase of the corresponding band, and $F$ deviates from $1$, indicating the breakdown of the adiabatic condition. The profile of $F$ is consistent with calculations using time-dependent perturbation (red line). Inset in (d) shows the variation of $F$ over a wider range of $\omega$, where in the large-$\omega$ limit, $F$ approaches unity again, as different Floquet bands are decoupled. For numerical calculations here, we take $\Omega=100$kHz and $\Delta_z=0$.}
  \label{fig:fig3}
\end{figure}

The break down of adiabatic condition can be systematically studied by plotting the Zak phase and the fidelity with increasing $\omega$. As shown in Fig.~\ref{fig:fig3}, both the Zak phase and fidelity deviates from their adiabatic values under larger $\omega$. While the oscillatory behavior in the Zak phase are due to Zitterbewegung-type interference~\cite{ZB,Zitterbewegung}, the overall decay is caused by the Landau-Zener tunneling, which can be well-explained using time-dependent perturbation. Specifically, under the time-dependent perturbation, the time-evolved state at time $T$ can be written as
\begin{align}
|\Psi_{\alpha}(T)\rangle& = e^{i(\varphi_{{\rm d},1}+\varphi_{{\rm z},1})}[|\mu^{\alpha}_{0,1}\rangle
+ \left.i\frac{\langle\mu^\alpha_{k(t),2}|\partial_t \mu^\alpha_{k(t),1}\rangle}{2|\epsilon_{k(t)}|}\right||\mu^\alpha_{0,2}\rangle]\nonumber\\
&=\frac{1}{\sqrt 2}e^{i(\varphi_{{\rm d},1}\pm\pi)}[ \left(\begin{matrix} 1\\ e^{\mp i 2\pi} \end{matrix}\right) \mp f(\omega,T)\left(\begin{matrix} -1\\ e^{\mp i 2\pi} \end{matrix}\right)],\label{eq:perturb}
\end{align}
where $f(\omega,t)=\frac{\Omega_{0}\Omega\cos(\omega t)+\Omega^2}{2[\Omega_{0}^2+2\Omega_{0}\Omega\cos(\omega t)+\Omega^2]^{3/2}}\omega$. As shown in Fig.~\ref{fig:fig3}(b), Eq.~(\ref{eq:perturb}) fits the overall decay profile of the fidelity.

When $\omega$ increases further, the perturbative calculation is no longer valid, while the fidelity increases again and approaches unity in the large-$\omega$ limit. This is because under the condition $\omega\gg \Omega,\Omega_0$, different Floquet bands are effectively decoupled from one another, such that the initial state is close to the eigenstate of the Floquet Hamiltonian $H^{\alpha}_{\rm F}$.

\begin{figure}[tbp]
  \centering
  \includegraphics[width=9cm]{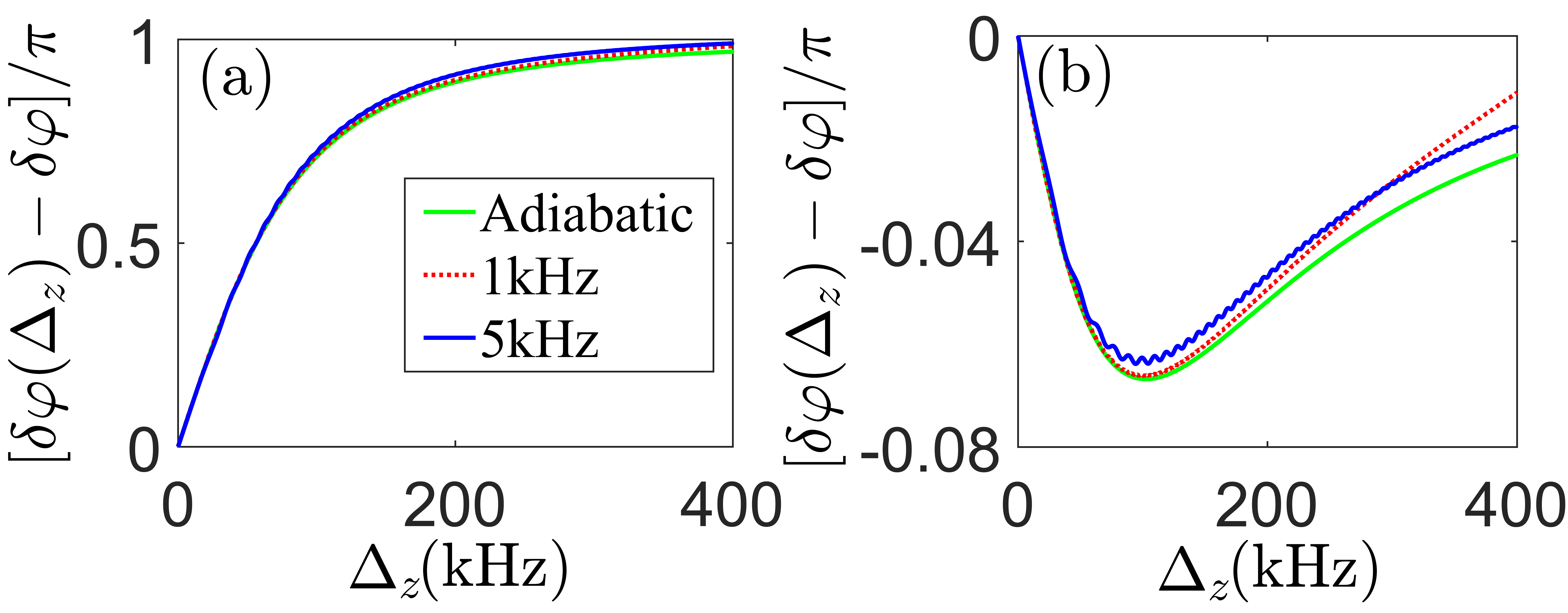}
  \caption{(a) Phase accumulated at different $\Delta_z$, with different $\omega$. We fix $\Omega=100$kHz and $\Omega_0=20$kHz, which generates topologically non-trivial bands at $\Delta_z=0$. (b) Phase accumulated at different $\Delta_z$, with different $\omega$. We fix $\Omega=100$kHz and $\Omega_0=180$kHz, which generates topologically trivial bands at $\Delta_z=0$. For both subplots, the blue solid and the red dashed curves  correspond to $\omega=1$kHz and $\omega=5$kHz, respectively. The solid green curves are the results in the adiabatic limit, calculated using Eq.~(\ref{eq:dz}). Since the adiabatic condition is satisfied, the green curves agree well with our numerical results.}
  \label{fig:fig4}
  \end{figure}

In previous discussions, we show that the Zak phase is quantized with small $\omega$ and a vanishing $\Delta_z$. When $\Delta_z$ becomes finite, the dynamics is mapped to the Bloch oscillation along a Rice-Mele lattice. The Zak phases of the corresponding bands are then no longer quantized, but continuously changes as $\Delta_z$ increases. As shown in Fig.~\ref{fig:fig4}(a), starting from a topologically non-trivial band, the Zak phase continuously changes from $\pi$ to $0$ with increasing $\Delta_z$. We note that a similar behavior has been observed in the Bloch oscillation of cold atoms in a superlattice~\cite{Bloch13}. In contrast, when starting from a topologically trivial band, the Zak phase remains close to $0$ [see Fig.~\ref{fig:fig4}(b)].

\section{Detection}

The Zak phase associated with the Bloch oscillation can be detected through interference measurements between different nuclear spin states in the clock-state manifold. As we have illustrated previously, due to the opposite sign of detunings in the laser coupling of different groups of clock states ($\alpha=\pm$), Zak phases accumulated in the same period of Bloch oscillation for different groups are also opposite in sign, whereas dynamic phases are the same. The phase difference between states in different groups after one period of Bloch oscillation thus reveal the accumulated Zak phase.
In practice, one can first initialize the atoms in an equal superposition of the clock states $\frac{1}{2}(|g,-\frac{1}{2}\rangle+|g,\frac{1}{2}\rangle+|e,-\frac{1}{2}\rangle+|e,\frac{1}{2}\rangle)$, before switching on the cross-coupling lasers for a duration $2\pi/\omega$. A standard interference measurement can then be applied to the states $|g,\pm \frac{1}{2}\rangle$ or $|e,\pm \frac{1}{2}\rangle$ to extract the Zak phase.




\section{Conclusion}

We show how periodic driving of the clock states in alkaline-earth(-like) atoms can simulate Bloch oscillations along a topological Wannier-Stark ladder in the synthetic Floquet dimension. The Zak phase associated with the topological lattice can be extracted from the Floquet dynamics, and is experimentally detectable through interference measurements between different nuclear spins of the clock-state manifold.
While our scheme can also be applied to generic two-level quantum systems such as vacancy centers in solids and superconducting qubits, the optical clock transition and the narrow line width of $^3P_0$ clock states make the system particularly suitable for the simulation of Floquet dynamics in the adiabatic regime. Specifically, since the line width of $^3P_0$ states is on the order of $10$mHz, it is convenient to make $\omega$ much smaller than $\Omega$ and $\Omega_0$.
Based on the current scheme, one may further consider spatially periodic coupling lasers, which offers the intriguing possibility of simulating complex band structures in higher dimensions.

\begin{acknowledgments}
We thank Wei Zheng for helpful comments. This work is is supported by  the National Natural Science Foundation of China (11974331), and the National Key Research and Development Program
of China (2016YFA0301700, 2017YFA0304100).
\end{acknowledgments}

\end{document}